\documentclass[runningheads]{llncs}
\usepackage[T1]{fontenc}
\usepackage{graphicx} 
\usepackage{algorithm}  
\usepackage[noend]{algpseudocode}
\usepackage{url}
\usepackage{listings}
\usepackage{xcolor}
\definecolor{codegreen}{RGB}{13,147,23}
\definecolor{commentgreen}{RGB}{2,112,10}
\definecolor{eminence}{RGB}{108,48,130}
\definecolor{weborange}{RGB}{255,165,0}
\definecolor{frenchplum}{RGB}{129,20,83}
\definecolor{skyblue}{RGB}{0,181,255}

\newcommand\+{\texttt{\raisebox{0.15ex}{+}}}

\title{Smart Fuzzing of 5G Wireless Software Implementation}
\author{Huan Wu\inst{1}
\and
Brian Fang\inst{2} \thanks{This work was done during an internship at Portland State University.}
\and
Fei Xie\inst{1}}
\authorrunning{H. Wu et al.}

\institute{Portland State University, Portland, OR 97201
\and
University of Pittsburgh, Pittsburgh, PA 15260}

\begin{document}\sloppy
\lstdefinelanguage{python}{
  morekeywords = [1]{from, import, def, if},
  morekeywords = [2]{},
  morekeywords = [3]{>,<,.,;,-,!,=,~, :=, <=, :|},
  keywordstyle = [1]\color{codegreen},
  keywordstyle = [2]\color{weborange},
  keywordstyle = [3]\color{weborange},
  sensitive = true,
  morecomment = [l]{//},
  morecomment = [s]{/*}{*/},
  morecomment = [s]{/**}{*/},
  commentstyle = \color{gray},
  morestring = [b]",
  morestring = [b]',
  stringstyle = \color{eminence}
}
\lstset{
  language={python},
  basicstyle={\footnotesize\ttfamily},
  identifierstyle={\footnotesize},
  commentstyle={\footnotesize\itshape},
  keywordstyle={\footnotesize\bfseries},
  ndkeywordstyle={\footnotesize},
  stringstyle={\footnotesize\ttfamily},
  frame={tb},
  breaklines=true,
  columns=[l]{fullflexible},
  showstringspaces=false
}

\maketitle

\begin{abstract}
In this paper, we introduce a comprehensive approach to bolstering the security, reliability, and comprehensibility of OpenAirInterface5G (OAI5G), an open-source software framework for the exploration, development, and testing of 5G wireless communication systems. Firstly, we employ AFL$\+\+$, a powerful fuzzing tool, to fuzzy-test OAI5G with respect to its configuration files rigorously. This extensive testing process helps identify errors, defects, and security vulnerabilities that may evade conventional testing methods. Secondly, we harness the capabilities of Large Language Models such as Google Bard to automatically decipher and document the meanings of parameters within the OAI5G codebase that are used in fuzzing. This automated parameter interpretation streamlines subsequent analyses and facilitates more informed decision-making. Together, these two techniques contribute to fortifying the OAI5G system, making it more robust, secure, and understandable for developers and analysts alike.
\end{abstract}

\section{Introduction}
\label{sec:introduction}
OpenAirInterface5G (OAI5G) \cite{oai5g} is an open-source framework designed to facilitate the exploration, development, and testing of 5G wireless communication technologies. This comprehensive platform fully implements the 5G protocol stack, encompassing critical radio access network components, including the physical layer, medium access control layer, radio resource control (RRC) layer, and core network elements. By harnessing the power of software-defined radios, OAI5G delivers flexibility for customization and real-world testing, empowering researchers, developers, and educators to immerse themselves in the intricacies of 5G networks. However, as the system's complexity grows, so do the potential vulnerabilities and defects that can undermine its reliability, stability, and security. Therefore, there is an urgent need for effective testing methods to detect and rectify these issues proactively.

Fuzzing \cite{fuzzing1}\cite{fuzzing2} is a software testing approach used to detect errors, defects, and security vulnerabilities in a software program or system by subjecting it to an extensive number of unexpected, malformed or random inputs. Leveraging fuzzing to evaluate OAI5G configuration files provides several advantages. It reveals vulnerabilities such as buffer overflows, memory leaks, and other security concerns triggered by unconventional or malformed configuration inputs. Additionally, it explores a broad spectrum of inputs, including edge cases and unconventional configurations, exposing issues that may manifest only under specific conditions. Fuzzing also automatically generates an extensive array of configuration inputs to ensure comprehensive coverage of possible scenarios and configurations, enhancing the OAI5G system's overall robustness. Among various fuzzing tools, AFL$\+\+$ \cite{AFL++} stands out as a top choice, being an enhanced version of the original American Fuzzy Lop (AFL) \cite{AFL} fuzzing tool. AFL$\+\+$ builds upon AFL's success with optimizations for faster test case generation and coverage measurement, diverse mutator strategies for wider test case variation, increased stability, and robustness.

Given the complex nature of the OAI5G codebase, which encompasses a multitude of parameters, some of these parameters can be challenging for developers and researchers to fully understand within the context of the 5G network framework. Fortunately, advancements in AI technology \cite{AIStudy} have opened new avenues for interpreting these parameters. Large Language models (LLMs), such as ChatGPT from OpenAI~\cite{ChatGPT} and Bard from Google AI \cite{Bard}, emerge as a promising solution. With advanced language capabilities and extensive training in text and code, LLMs excel in interpreting OAI5G parameters. Leveraging their natural language understanding, LLMs elucidate complex parameter meanings, enhancing accessibility and comprehension for OAI5G developers and researchers. 

This paper presents a comprehensive approach that leverages advanced testing techniques and AI-driven tools to enhance the OAI5G system's security, reliability, and comprehensibility, ultimately contributing to its robustness and accessibility. It presents two essential techniques: leveraging AFL$\+\+$ for comprehensively fuzzing OAI5G's configuration files and harnessing Google Bard's API for automated parameter interpretation within the OAI5G codebase. These methods collectively enhance OAI5G's resilience, making it more secure and user-friendly, and have significant potential in the broader context of 5G network development.

The remainder of this paper is organized as follows. Section 2 provides an overview of the 5G New Radio and AFL$\+\+$. Section 3 outlines the methodology of our approach. Section 4 discusses the implementation details of our approach. Finally, Section 5 concludes the paper and outlines future research directions.

\section{Background}
\label{sec:Background}
\subsection{5G New Radio}
The 3GPP (3rd Generation Partnership Project) 5G New Radio (NR) architecture \cite{ETSI} encompasses both the radio access network and the core network, each with distinct functionalities. Within the 5G NR radio access network, various key components collaborate to provide wireless connectivity to user devices. At the heart of the RAN lies the Next-Generation NodeB (gNB), tasked with radio resource control, physical layer management, and radio protocol handling. The gNB establishes communication with user equipment (UE) and other gNBs as necessary. On the other hand, the 5G NR core network is designed for flexibility, scalability, and versatile support for numerous services and applications. It adopts service-based architecture principles, enabling capabilities such as network slicing and efficient resource allocation.

\begin{figure}[ht]
\centering
\includegraphics[width=4.5in]{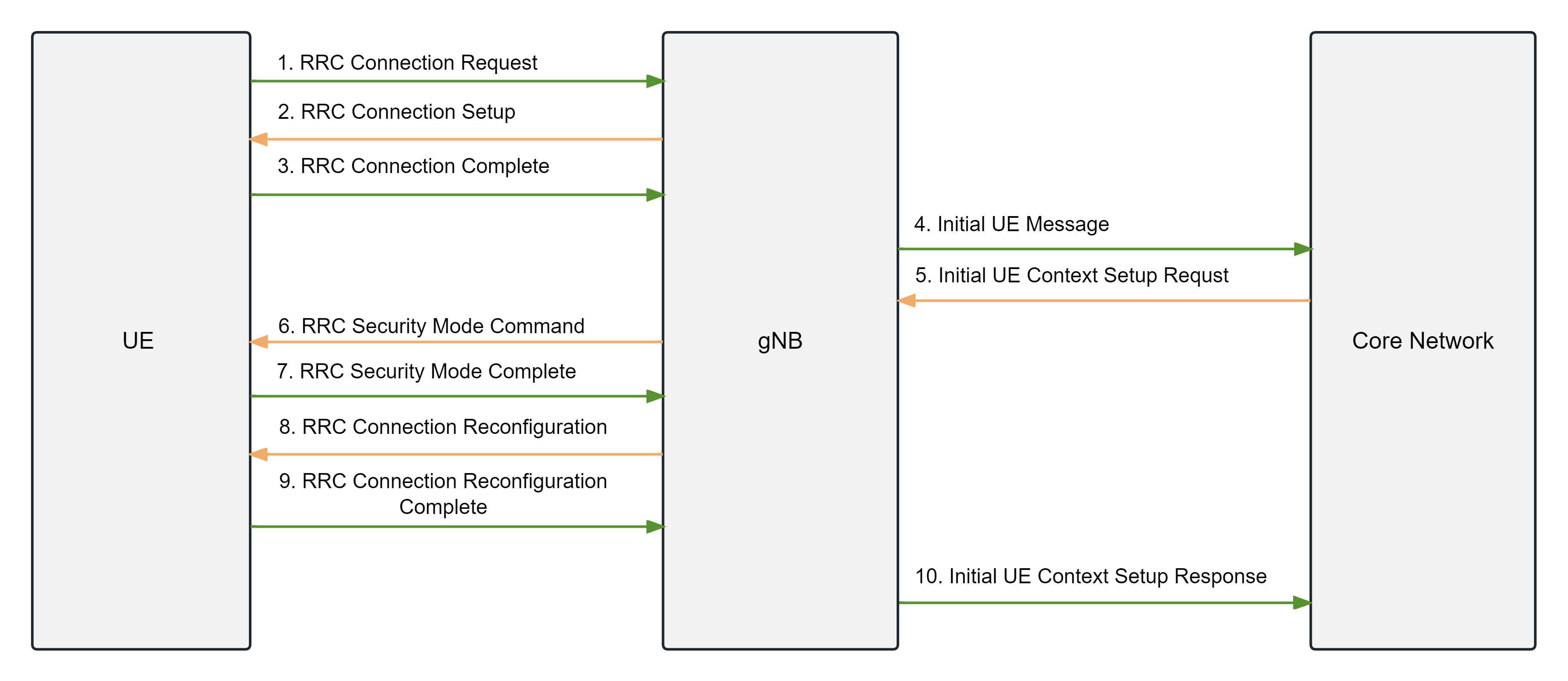}
\caption{5G NR Network Connection Process}
\label{fig:op1}
\end{figure}

The communication process in a 5G NR network, involving the UE, the gNB, and the core network, encompasses several steps for connection establishment and maintenance. As depicted in Figure~\ref{fig:op1}, this process unfolds as follows: (1) RRC Connection Request: The initiation phase involves the UE dispatching a request to the gNB, signaling its intent to establish communication. (2) RRC Connection Setup: In response to the request, the gNB configures communication parameters. 
(3) RRC Connection Setup Complete: The UE confirms the successful setup.
(4) Initial UE Message: The gNB informs the core network, initiating subsequent actions. (5) Initial UE Context Setup Request: The core network initiates preparations with a request to the gNB. (6) RRC Security Mode Command: The gNB sends a security mode command to the UE to ensure security. (7) RRC Security Mode Complete: The UE acknowledges security readiness. (8) RRC Connection Reconfiguration: The gNB adjusts communication as needed. (9) RRC Connection Reconfiguration Complete: The UE confirms seamless adjustments. (10) Initial UE Context Setup Response: The gNB finalizes the UE's context within the network, ensuring a stable and secure connection.

In the communication process of a 5G NR network, the gNB's role in configuring communication parameters is crucial for establishing a robust and efficient connection. These parameters encompass vital communication aspects like frequency bands, modulation schemes, and coding rates. Proper configuration by the gNB ensures optimized performance and reliability in the communication between the UE and the core network. 

\subsection{AFL$\+\+$}
AFL$\+\+$ is a powerful and indispensable fuzzing tool in software security and vulnerability discovery. It builds upon AFL by incorporating cutting-edge research and enhancements, extending its capabilities with additional power schedules, and offering an evolving API that eliminates the need for forking or patching. It features the Custom Mutator API for customized fuzzing processes and target-specific mutators. Supporting multiple instrumentation backends (LLVM, GCC, QEMU, Unicorn, QBDI), it provides a flexible proxy module for forwarding test cases. Compatible with various operating systems and distributions, AFL$\+\+$ includes optimizations like a Linux Kernel Module inspired by Perffuzz \cite{xu2017designing}, significantly boosting performance, especially in parallel fuzzing, without requiring target program recompilation.

\section{Method}
\label{sec:method}
In this section, we outline the core techniques that constitute our approach to fortifying the OAI5G system.
These techniques cover comprehensive fuzzing of OAI5G's configuration files and the automated interpretation of parameters within the OAI5G codebase. The following details provide essential insights into bolstering the overall robustness of OAI5G.

\subsection{Fuzzing With AFL$\+\+$}
\begin{figure}[ht]
\centering
\includegraphics[width=2.8in]{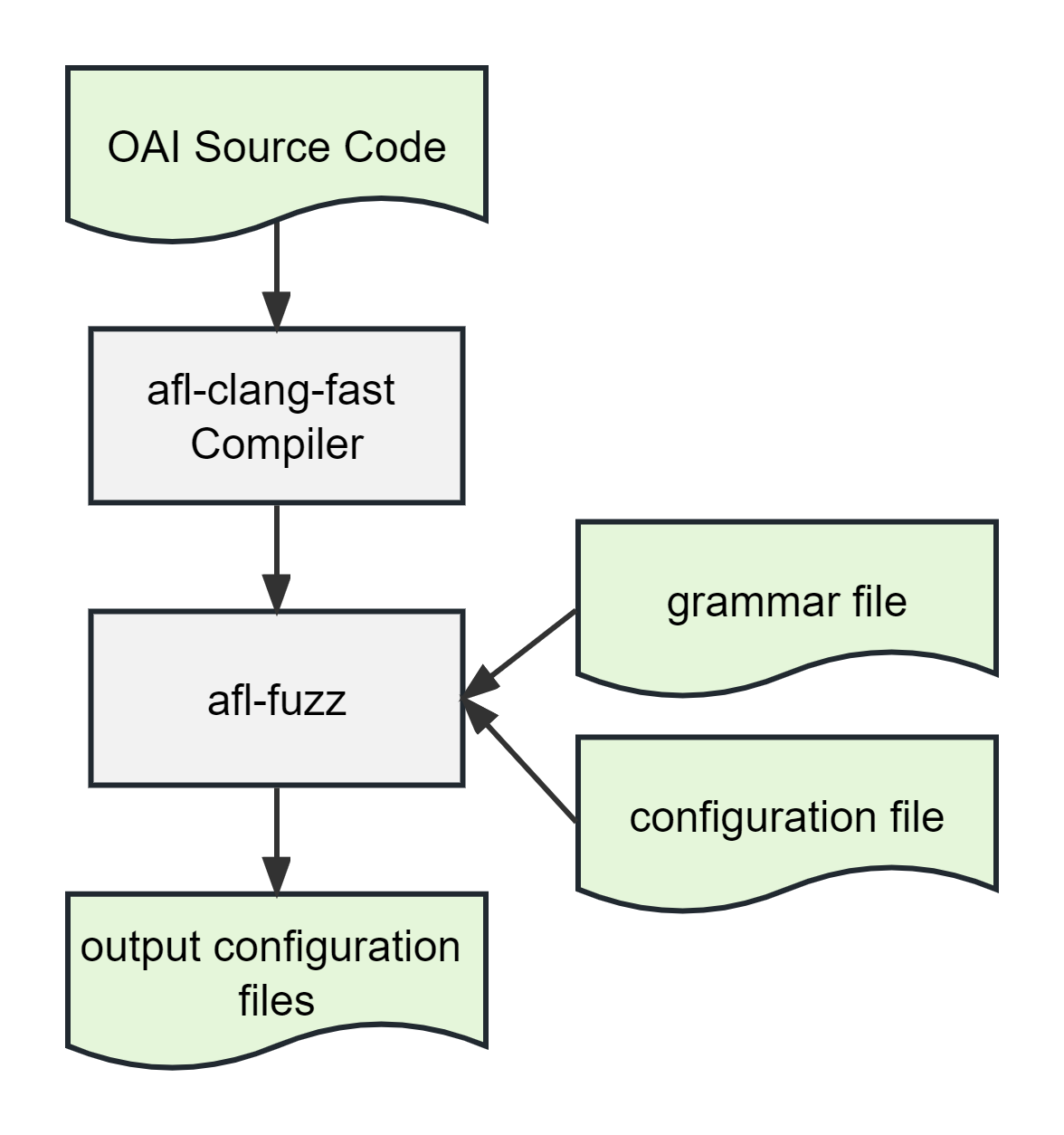}
\caption{AFL$\+\+$ Testing Procedure}
\label{fig:afl1}
\end{figure}

This section demonstrates the systematic approach employed to leverage AFL$\+\+$ for the fuzzing of OAI5G. Figure~\ref{fig:afl1} provides a comprehensive overview of this approach.
It starts with a critical adaptation: the substitution of the conventional GCC compiler in the OAI5G codebase with AFL$\+\+$'s afl-clang-fast compiler, primarily for instrumentation purposes.
This choice is grounded in several reasons. Firstly, afl-clang-fast is based on the LLVM infrastructure, known for its reliability and widespread usage in compiler technology. Additionally, it offers a notable speed advantage over alternative compilers, making it highly efficient for large-scale codebases such as OAI5G. These efficiencies translate into reduced overhead during instrumentation. Furthermore, afl-clang-fast incorporates optimizations tailored for precise code instrumentation, enhancing our fuzz testing capabilities.

Following the instrumentation of the OAI5G target, the subsequent significant step entails the strategy implementation of AFL$\+\+$'s afl-fuzz. 
This implementation is geared toward conducting comprehensive fuzzing of OAI5G by exploring its configuration parameters. This phase is specifically dedicated to systematically assessing the parameters within the structured configuration file used as input by the gNB. To handle the structures of the configuration file, we craft a corresponding grammar file. AFL$\+\+$ introduces a powerful grammar mutator inspired by the principles of the F1 fuzzer \cite{f1fuzzer} and Nautilus \cite{Nautilus}, significantly enhancing its ability to manipulate structured input formats like JSON and Ruby. This augmentation makes AFL$\+\+$ a versatile tool for thorough and methodical fuzzing. By generating a diverse spectrum of input variations, we systematically scrutinize the configuration files for potential vulnerabilities, unique scenarios, and unforeseen anomalies. 

\subsection{Automated Parameter Interpretation}
Leveraging Bard's potential to automatically generate interpretations for OAI5G codebase parameters is another essential component for our smart fuzzing. We develop an algorithm to preprocess input parameters and facilitate the automated interpretation of complex ones.

This algorithm, showcased in Algorithm 1, demonstrates the extraction and analysis of test case parameters, their variable names, and their meanings within the OAI5G source code. It starts by parsing an input file to identify test cases and storing them in an ArrayList. Subsequently, it processes these test cases to isolate unique parameters and their associated value ranges, managing this data within a LinkedHashMap. Following this, it proceeds to locate the corresponding variable names for these parameters, employs the Bard API to retrieve their meanings within the OAI5G source code, and constructs result strings that encompass parameter details, variable names, and meanings. Lastly, these results are systematically written to an output file.

By seamlessly combining Algorithm 1 with our comprehensive fuzzing approach, we streamline and enhance the procedure for automating the generation of meanings for OAI5G parameters. This tool empowers us to better understand these elements within the OAI5G codebase, associating them seamlessly with our fuzzing strategies and fortifying the OAI5G system.
\begin{algorithm}[H]
\centering    
    \caption{\small AutoExplainParams(input, output)}
        \begin{algorithmic}[1]
            \State tests \texttt{=} ArrayList()
            \For{line \textbf{in} readLines(input)}
                \If{isTest(line)}
                    \State tests.add(line)
                \EndIf            
            \EndFor
            \State paramsMap \texttt{=} LinkedHashMap()
            \For{test \textbf{in} tests}
                \State uniqueParams \texttt{=} extractUniqueParams(test)
                \For{param \textbf{in} uniqueParams}
                    \State params.put(param, getParamRange(param)
                \EndFor
            \EndFor
            \State results \texttt{=} ArrayList()
            \For{param \textbf{in} params}            
                \State varName \texttt{=} findParamName(param)
                \State varMeaning \texttt{=} callingBardAPI(varName)
                \State result \texttt{=} "-" \texttt{+} param \texttt{+} " (" \texttt{+} varName \texttt{+} ") \texttt{->} " \texttt{+} varMeaning
                \State results.add(result)
            \EndFor
            \State writeLines(output, results)            
        \end{algorithmic}
\end{algorithm}

\section{Implementation}
\label{sec:implementation}
This section expands on the implementation details of our approach. It covers the execution of comprehensive fuzzing using AFL$\+\+$ and the integration of automated parameter explanation techniques for OAI5G parameters.

\subsection{Execution of Fuzzing}
We first discuss the implementation of fuzz testing with AFL$\+\+$ using OAI5G's configuration files. We initiate the fuzz testing process using a sample configuration file from OAI5G as the seed input, as depicted in Figure~\ref{fig:afl_configure}.
\begin{figure}[htb!]
    \centering
\begin{lstlisting}[language=Python]
gNBs = ({
......
do_CSIRS = 1;
do_SRS = 1;
pdcch_ConfigSIB1 = ({
controlResourceSetZero = 12;
searchSpaceZero = 0;});
servingCellConfigCommon = ({
......
absoluteFrequencySSB = 641280;
dl_frequencyBand = 78;
dl_absoluteFrequencyPointA = 640008;
dl_offstToCarrier = 0;
dl_carrierBandwidth = 106;
......})
......})
\end{lstlisting}
    \caption{Sample Configuration File}
    \label{fig:afl_configure}
\end{figure}
 This figure shows the extraction of select parameters and their respective values from the configuration file. To align with AFL$\+\+$'s grammar mutator, we provide a corresponding grammar file for the configuration file, represented in Figure~\ref{fig:afl_grammar}. This JSON-formatted grammar file comprises key-value pairs, where each key includes a grammar token enclosed within angle brackets and the associated value consisting of grammar rules in the form of string lists.
 \begin{figure}[htb!]
    \centering
\begin{lstlisting}[language=Python]
{......
"<26>": [["do_CSIRS = ", "<27>", ";\n"]],
"<27>": [["1"], ["0"]],
"<28>": [["do_SRS = ", "<29>", ";\n"]],
"<29>": [["1"], ["0"]],
"<30>": [["pdcch_ConfigSIB1 = (\n{\n", "<31>", "<33>", "}\n);\n"]],
"<31>": [["controlResourceSetZero = ", "<32>", ";\n"]],
"<32>": [["3"],["4"],["5"],["6"],["7"],["8"],["9"],["10"],["11"],["12"]],
"<33>": [["searchSpaceZero = ", "<34>", ";\n"]],
"<34>": [["0"],["1"],["2"],["3"],["4"],["5"],["6"],["7"],["8"],["9"]],
......
"<38>": [["absoluteFrequencySSB = ", "<39>", ";\n"]],
"<39>": [["641280"],["641272"],["433096"],["642016"],["623232"]],
"<40>": [["dl_frequencyBand = ", "<41>", ";\n"]],
"<41>": [["78"],["41"],["66"],["257"],["261"]],
"<42>": [["dl_absoluteFrequencyPointA = ", "<43>", ";\n"]],
"<43>": [["640008"],["43000"],["623336"],["640032"],["620016"]],
......
"<48>": [["dl_carrierBandwidth = ", "<49>", ";\n"]],
"<49>": [["24"],["25"],["66"],["106"],["133"],["162"],["217"],["273"]],
......}
\end{lstlisting}
    \caption{Sample Grammar File}
    \label{fig:afl_grammar}
\end{figure}
 These rules can represent either concrete strings or references to other grammar tokens. Leveraging this provided grammar, the grammar mutator constructs a tree-like structure for each input test case, which is subsequently translated into a concrete input format compatible with the target application. 

Given the extensive scale of the OAI5G codebase, a strategic consideration pertains to adjusting the timeout period to "10000$\+$" milliseconds for the forthcoming implementation of AFL$\+\+$'s afl-fuzz. This elongated timeout duration accounts for the intricate nature of the OAI5G codebase and aligns to scrutinize the configuration parameters embedded within the configuration file thoroughly. By extending the timeout to this duration, we aim to ensure that the fuzzing process extensively explores many potential code paths and test scenarios, enabling the effective identification of vulnerabilities, exceptional conditions, and unanticipated behaviors within the OAI5G ecosystem.
\begin{table}[ht]
    \centering
    \caption{Sample Crash Test Cases}
    \label{tab:crashes}
    \begin{tabular}{| l | c | c | c | c | c |c | }
     \hline
     Parameters & Initial  & Case1 & Case2 & Case3 & Case4 & Case5 \\
     \hline
     do$\_$CSIRS & 1  & 0 & 0 & 0 & 0 & 1 \\
     \hline
      do$\_$SRS & 1  & 0 & 0 & 0 & 1 & 1 \\
     \hline
     controlResourceSetZero & 12  & 9 & 3 & 9 & 6 & 12 \\
     \hline
      searchSpaceZero & 0  & 9 & 8 & 9 & 8 & 0 \\
     \hline
      absoluteFrequencySSB & 641280  & 433096 & 641272 & 642016 & 623232 & 641280 \\
     \hline
      dl$\_$frequencyBand & 78  & 78 & 78 & 41 & 78 & 257 \\
     \hline
      dl$\_$absoluteFrequencyPointA & 640008  & 640008 & 43000 & 43000 & 43000 & 640008 \\
     \hline
     dl$\_$carrierBandwidth & 106  & 106 & 25 & 25 & 24 & 106 \\
     \hline
    \end{tabular}
    \label{samplescrashes}
\end{table}

Through the fuzzing process, several crash cases are identified, and a selection of these cases is presented in Table~\ref{samplescrashes}. Notably, common patterns among these crashes illuminate specific inputs or configurations prone to issues, guiding focused improvement efforts. Additionally, evaluating the severity and impact of each crash has enabled a more efficient prioritization of debugging and enhancement tasks. These cases also serve as valuable practical examples for documentation and reference, fostering improved comprehension of problems and their solutions.

\subsection{Auto-Explaining Parameters Integration}
As an example, we employ the physical layer test cases and parameters embedded within the OAI5G codebase to exemplify the utilization of Bard for automated parameter interpretation.
\begin{figure}[htb!]
    \centering
\begin{lstlisting}[language=Python]
Description = nr_pbchsim Test cases.  
                (Test1: PBCH-only, 106 PRB),
                (Test2: PBCH and synchronization, 106PBR),
                (Test3: PBCH-only, 217 PRB),
                (Test4: PBCH and synchronization, 217 RPB),
                (Test5: PBCH-only, 273 PRB),
                (Test6: PBCH and synchronization, 273 PRB),
                (Test7: PBCH and synchronization, 106PBR, SSB SC OFFSET 6)
main_exec_args = -s-11 -S-8 -n10 -R106
                -s-11 -S-8 -n10 -o8000 -I -R106
                -s-11 -S-8 -n10 -R217
                -s-11 -S-8 -n10 -o8000 -I -R217
                -s-11 -S-8 -n10 -R273
                -s-11 -S-8 -n10 -o8000 -I -R273
                -s-11 -S-8 -n10 -R106 -O6
\end{lstlisting}
    \caption{Sample Input File}
    \label{fig:bard_input}
\end{figure}
Figure~\ref{fig:bard_input} presents a snapshot of the input file extracted from the log generated during the execution of the PBCH-related physical layer test cases in OAI5G's autotests. This example encompasses seven distinct use cases, each featuring seven unique sets of input parameter combinations. We aim for Bard to explain each parameter's meaning in this context.

\begin{figure}[ht]
\centering
\includegraphics[width=4.2 in]{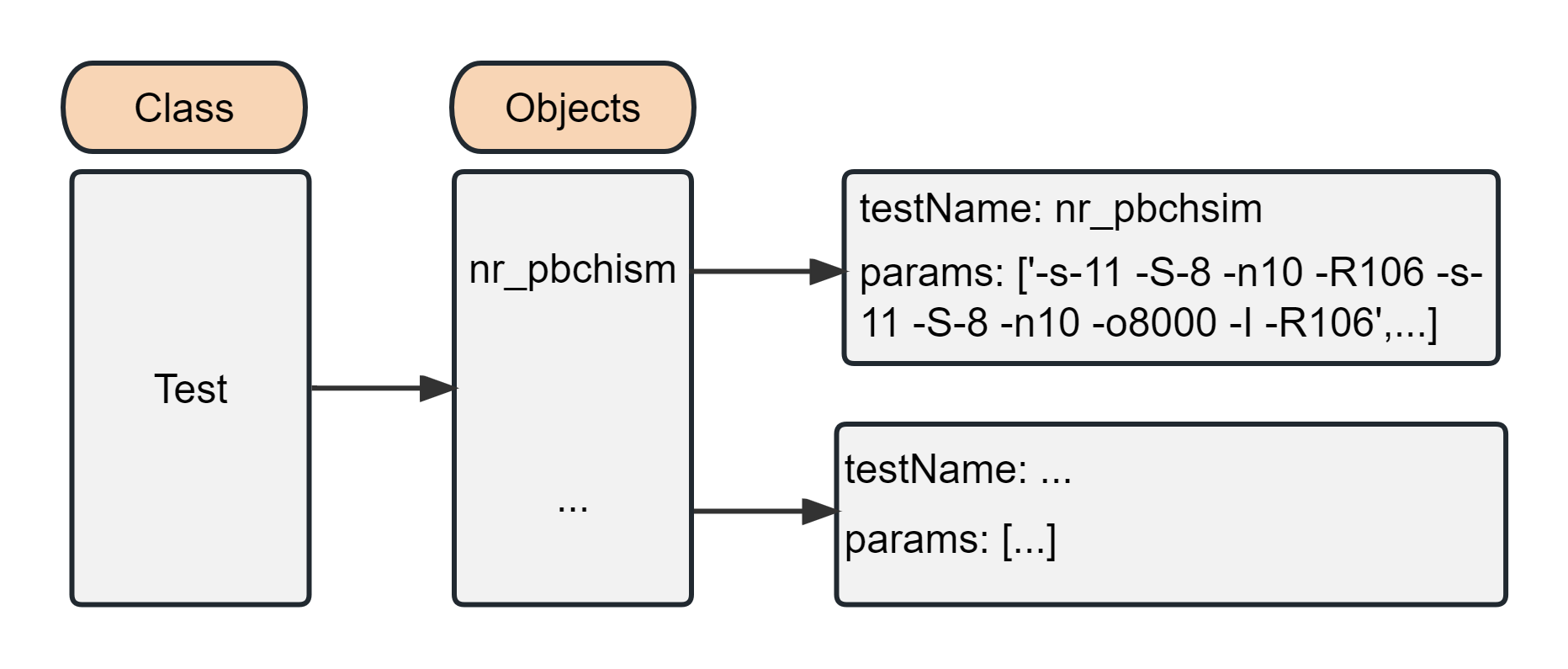}
\caption{Test Case Storage}
\label{fig:bard_test_case_storage}
\end{figure}

As outlined in Algorithm 1 within Section 2.3, Figure~\ref{fig:bard_test_case_storage} illustrates the process wherein each test case is organized as a Test Object, preserving their respective test names and parameters as inherent properties. 
It proceeds to iterate through this ArrayList, establishing a LinkedHashMap for every test case to create a mapping between parameter names and a LinkedHashSet encompassing parameter values. For each parameter, a search operation is performed within the corresponding test case file to recover the associated variable name.
\begin{figure}[ht]
\centering
\includegraphics[width=4 in]{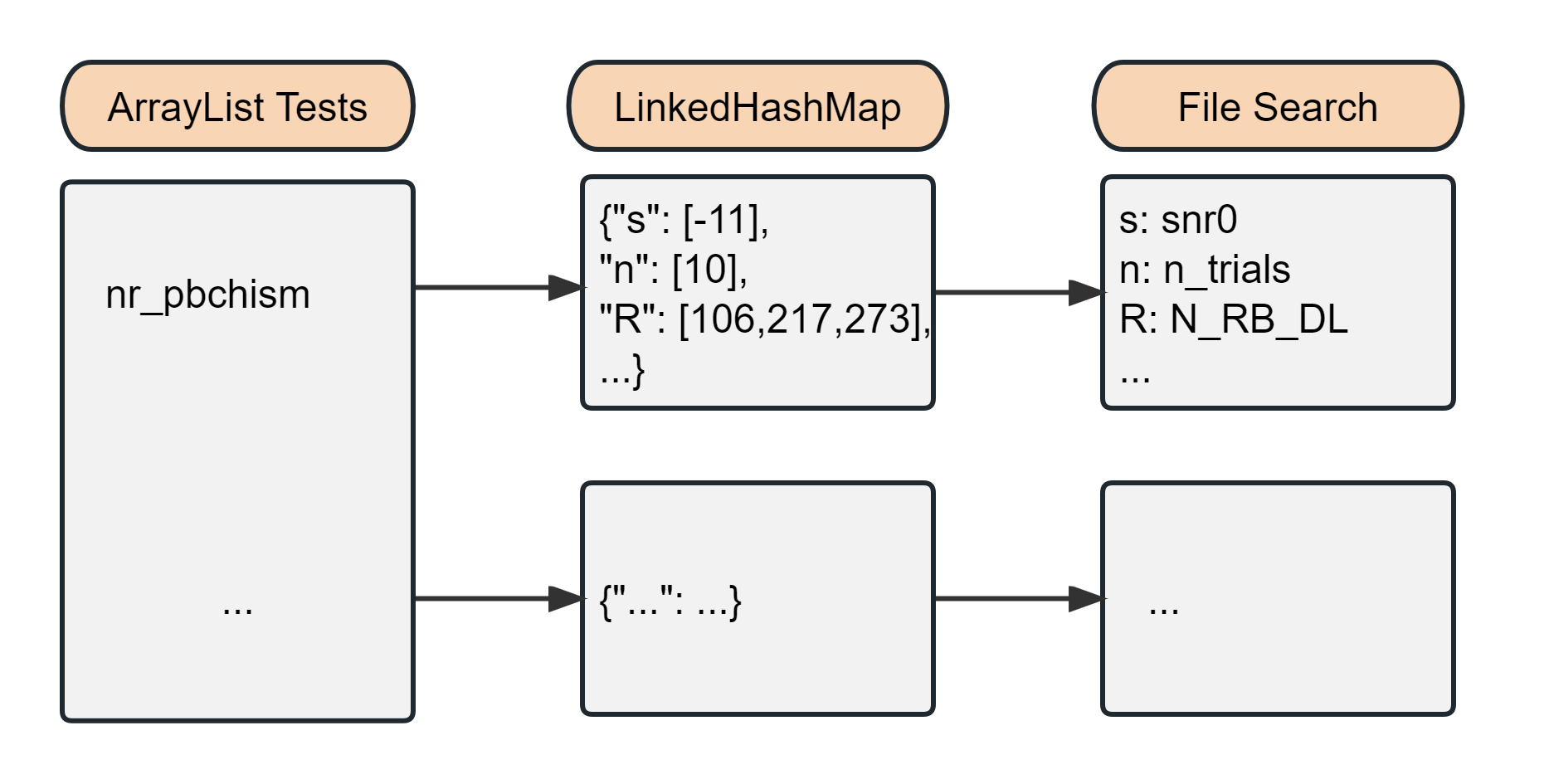}
\caption{Test Case Structure}
\label{fig:bard_test_case_structure}
\end{figure}
Figure~\ref{fig:bard_test_case_structure} further elucidates the architecture of individual test cases within the ArrayList. Each case is structured as a LinkedHashMap, featuring a parameter name as a key, complemented by a LinkedHashSet that encapsulates the associated range of values. The algorithm subsequently queries the relevant files to determine the complete variable name corresponding to each parameter.

With these preparations complete, the algorithm leverages Bard's API to access the definitions of these variable names within OAI5G. Finally, it generates an output file with crucial information such as the test case name, variable names, meanings, and the respective range of values. These details are in a structured format, exemplified in Figure~\ref{fig:bard_output}, which displays this comprehensive information for the PBCH-related test case.
\begin{figure}[htb!]
    \centering
\begin{lstlisting}[language=Python]
nr_pbchsim
-s (snr0) -> Initial Signal-to-Noise Ratio (SNR) for PBCH (Physical Broadcast Channel) simulation: -11
-S (snr1) -> Additional Signal-to-Noise Ratio (SNR) for PBCH simulation: -8
-n (n_trials) -> Number of PBCH simulation trials: 10
-R (N_RB_DL) -> Number of resource blocks for downlink: 106, 217, 273
-o (cfo) -> Carrier Frequency Offset (CFO) value for PBCH simulation: 8000
-I (run_initial_sync) -> Run initial synchronization
-O (ssb_subcarrier_offset) -> Subcarrier offset for Synchronization Signal Blocks (SSBs): 6
\end{lstlisting}
    \caption{Test Case Output}
    \label{fig:bard_output}
\end{figure}

The described systematic process, facilitated by Bard's API, clarifies the significance of vital parameters in the OAI5G codebase and simplifies follow-up analyses. This understanding empowers developers and analysts to make informed decisions, uncover optimizations, and bolster the overall efficiency and reliability of the OAI5G system.


\section{Conclusions and Future Work}
\label{sec:conclusion}
This study presents a holistic approach to enhance the security, reliability, and comprehensibility of the OAI5G system through smart fuzzing. Two core techniques underpin this strategy. Firstly, we employ AFL$\+\+$ to meticulously evaluate OAI5G's configuration files. This rigorous testing reveals vulnerabilities, defects, and security weaknesses often missed by conventional methods. Secondly, Google Bard's API automates the interpretation and documentation of parameters in the OAI5G codebase, streamlining subsequent analyses and enabling informed decisions. These combined efforts strengthen the OAI5G system, benefiting developers and analysts alike. Looking ahead, avenues for research and development include extending our fuzzing approach to critical components, like parameters influencing the communication between gNB and UE. Leveraging AI technologies like Google Bard and ChatGPT also holds promise, offering insights, recommendations, and automation for enhanced testing and analysis.

\newpage
\bibliographystyle{splncs04}
\bibliography{ref.bib}

\end{document}